\documentclass[twocolumn,showpacs,preprintnumbers,amsmath,amssymb,superscriptaddress,floatfix,nofootinbib]{revtex4}

\usepackage{graphicx}
\usepackage{amsmath}
\usepackage{amsfonts}
\usepackage{CJK}
\usepackage{float}

\begin{document}

\title{Prediction of possible exotic states in the $\eta \bar{K}K^*$ system}

\author{Xu Zhang}
\email{zhangxu@impcas.ac.cn} \affiliation{Institute of Modern
Physics, Chinese Academy of Sciences, Lanzhou 730000, China}
\affiliation{School of Nuclear Science and Technology, University of
Chinese Academy of Sciences, Beijing 101408, China}

\author{Ju-Jun Xie}
\email{xiejujun@impcas.ac.cn} \affiliation{Institute of Modern
Physics, Chinese Academy of Sciences, Lanzhou 730000, China}
\affiliation{School of Nuclear Science and Technology, University of
Chinese Academy of Sciences, Beijing 101408, China}
\affiliation{School of Physics, Zhengzhou University, Zhengzhou,
Henan 450001, China}

\begin{abstract}

We investigate the $\eta \bar{K}K^*$ three body system in order to
look for possible $I^G(J^{PC})=0^+(1^{-+})$ exotic states in the
framework of the fixed center approximation to the Faddeev equation.
We assume the scattering of $\eta$ on a clusterized system
$\bar{K}K^*$, which is known to generate the $f_1(1285)$, or a
$\bar{K}$ on a clusterized system $\eta K^*$, which is shown to
generate the $K_1(1270)$. In the case of the
$\eta$-$(\bar{K}K^*)_{f_1(1285)}$ scattering, we find evidence of a
bound state $I^G(J^{PC})=0^+(1^{-+})$ below the $\eta{f_1(1285)}$
threshold with mass around 1700 MeV and width about 180 MeV.
Considering the $\bar{K}$-$(\eta K^*)_{K_1(1270)}$ scattering, we
obtain a bound state $I(J^{P})=0(1^{-})$ just below the
$\bar{K}{K_1(1270)}$ threshold with a mass around 1680 MeV and a
width about 160 MeV.
\end{abstract}

\date{\today}

\pacs{  } \maketitle


\section{Introduction}

Exotic states cannot be described by the traditional quark model and
may be of more complex structure allowed in QCD such as glueballs,
hybrid mesons and multiquark states. The discovery of exotic states
and the study of their structure will apparently extend our
knowledge of the strong interaction
dynamics~\cite{Liu:2019zoy,Guo:2017jvc,Chen:2016spr}.

A meson with quantum numbers $J^{PC}=1^{-+}$ which is excluded by
the traditional quark model with $q \bar{q}$ picture is an exotic
state \cite{Amsler:2004ps}. Interestingly, three isovector
$J^{PC}=1^{-+}$ exotic candidates, namely $\pi_1(1400)$,
$\pi_1(1600)$, and $\pi_1(2015)$ have been reported
experimentally~\cite{Tanabashi:2018oca}. On the theoretical side,
the isovector exotic states are interpreted as hybrid mesons within
a few different theoretical approaches such as the flux tube
model~\cite{Isgur:1984bm,Close:1994hc,Page:1998gz}, ADS/QCD
model~\cite{Kim:2008qh,Bellantuono:2014lra}, and Lattice
QCD~\cite{Dudek:2010wm,Dudek:2009qf,Bernard:2003jd}. In addition,
the hybrid meson decay properties are studied within the framework
of QCD sum rules in
Refs.~\cite{Huang:2017pzh,Huang:2016upt,Huang:2014hya,Chen:2010ic}.
Some works suggest that the isovector exotic state might be a
fourquark state~\cite{Chen:2008qw} or a molecule/four-quark mixing
state~\cite{Narison:2009vj}. On the other hand, the three body
system can also carry the quantum numbers $J^{PC}=1^{-+}$. In
Ref.~\cite{Zhang:2016bmy}, by keeping the strong interactions of
$\bar{K}K^*$ which generate the $f_1(1285)$
resonance~\cite{Roca:2005nm,Lutz:2003fm}, the $\pi \bar{K}K^*$ three
body system was investigated within the framework of the fixed
center approximation (FCA) to the Faddeev equation, where the
$\pi_1(1600)$ could be interpreted as a dynamically generated state
from $\pi$-$(\bar{K}K^*)_{f_1(1285)}$ system.

In principle, an isoscalar exotic state is also possible, though not
observed experimentally~\cite{Page:1998gz,Bernard:2003jd}. In fact,
these isoscalar exotic states were studied with the QCD sum rules
analysis using the tetraquark currents~\cite{Chen:2008ne}, where the
obtained mass is around $1.8 \sim 2.1$ GeV, and the decay width is
about 150 MeV.

In this paper we study the $\eta \bar{K}K^*$ three body system in
order to look for possible $I^G(J^{PC}) = 0^+(1^{-+})$ exotic states
within the FCA approach, which has been used to investigate the
interaction of $K^-d$ at
threshold~\cite{Gal:2006cw,Barrett:1999cw,Kamalov:2000iy}. Within
the FCA approach, a possible state in three body system $K^- pp$
according to the calculation done within the framework of FCA
approach~\cite{Bayar:2011qj,Bayar:2012hn}, has been supported by the
J-PARC experiments~\cite{Ajimura:2018iyx}. In Ref.~\cite{Xie:2011uw}
the $\Delta_{5/2^+}(2000)$ puzzle is solved in the study of the
$\pi$-$(\Delta \rho)$ interaction, and in Ref.~\cite{Xie:2010ig}, a peak is found around 1920 MeV indicating a $NK \bar{K}$ state with
$I = 1/2$ around that energy, which supports the existence of a $N^*$
resonance with $J^P = 1/2^+$ around 1920 MeV obtained in
Refs.~\cite{Jido:2008kp,MartinezTorres:2009cw,MartinezTorres:2008kh,MartinezTorres:2010zv},
where the full Faddeev calculations were done. Recently, the
predictions of several heavy flavor resonance states in three body
system have been carried out within the framework of FCA approach
like $\bar{K}^{(*)} B^{(*)}\bar{B}^{(*)}$ \cite{Ren:2018qhr},
$D^{(*)}B^{(*)}\bar{B}^{(*)}$ \cite{Dias:2018iuy}, $\rho B^*
\bar{B}^*$~\cite{Bayar:2015zba}, $\rho D^*
\bar{D}^*$~\cite{Bayar:2015oea,Xiao:2012dw}, $DKK$
$(DK\bar{K})$~\cite{Debastiani:2017vhv}, $BDD$
$(BD\bar{D})$~\cite{Dias:2017miz}, and
$K\bar{D}D^*$~\cite{Ren:2018pcd}. While the $DDK$ system is
investigated in coupled channels by solving the Faddeev equations
using the two-body inputs in Ref.~\cite{MartinezTorres:2018zbl},
where it is found that an isospin $1/2$ state is formed at 4140 MeV
when the $D^*_{s0}(2317)$ is generated in the $DK$ subsystems. And,
such a result is compatible with the one found in
Ref.~\cite{SanchezSanchez:2017xtl} where the system
$D$-$D^*_{s0}(2317)$ was studied without considering explicit
three-body dynamics. In a more recent work~\cite{Wu:2019vsy}, where
the Gaussian expansion method was used, the existence of these $DDK$
states has been further confirmed. These above examples show that
the results of the FCA prove to be rather reasonable. However, as
important as it is to know the success of the FCA, there are
problems for the FCA to study the $\phi K \bar{K}$
system~\cite{MartinezTorres:2010ax} (more details about the limits to the FCA can be found there), for which the $\phi(2175)$ can
be reproduced by full Faddeev equation calculations with the $\phi K
\bar{K}$ system~\cite{MartinezTorres:2008gy}.

There are two possible scattering cases for the $\eta \bar{K}K^*$
three body system since the $\bar{K}K^*$ and $\eta K^*$ system lead
to the formation of two dynamically generated resonances,
$f_1(1285)$ and $K_1(1270)$, respectively. Based on the two body
$\eta \bar{K}$, $\eta K^*$, and $\bar{K}K^*$ scattering amplitudes
obtained from the chiral unitary
approach~\cite{Roca:2005nm,Geng:2006yb,Guo:2005wp}, we perform an
analysis of the $\eta$-$(\bar{K}K^*)_{f_1(1285)}$ and
$\bar{K}$-$(\eta K^*)_{K_1(1270)}$ scattering amplitude, which will
allow us to predict the possible exotic states with quantum numbers
$I^G(J^{PC})=0^+(1^{-+})$.

This paper is organized as follows. In Sec.~II we present the FCA
formalism and ingredients to analyze the
$\eta$-$(\bar{K}K^*)_{f_1(1285)}$ and $\bar{K}$-$(\eta
K^*)_{K_1(1270)}$ systems. In Sec.~III, numerical results and
discussions are shown. Finally, a short summary is given in Sec.~IV.

\section{Formalism and ingredients} \label{sec:formalism}

\subsection{Fixed center approximation formalism}

\begin{figure*}[htbp]
\begin{center}
\includegraphics [scale=0.6] {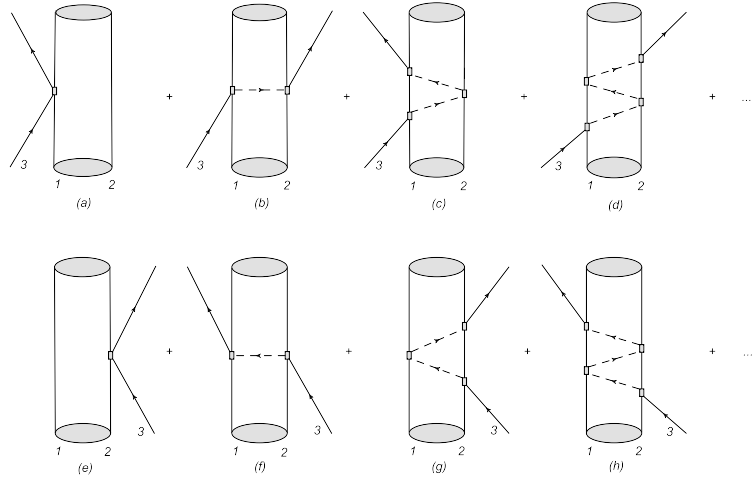}
\caption{Diagrammatic representation of the FCA to Faddeev
equations.} \label{fig:fca-diagram}
\end{center}
\end{figure*}

Within the framework of FCA, we consider $\bar{K}K^*(\eta K^*)$ as a
cluster and $\eta(\bar{K})$ interacts with the components of the
cluster. The total three body scattering amplitude $T$ can be
simplified as the summation of the two partition functions $T_1$ and
$T_2$, summing for all the diagrams of Fig.~\ref{fig:fca-diagram}
starting with the interaction of particle 3 with particle 1(2) of
the cluster. Then the FCA equations can be written in terms of $T_1$
and $T_2$, which sum up to the total scattering amplitude $T$, and
read~\cite{Barrett:1999cw,Deloff:1999gc,Kamalov:2000iy}
\begin{eqnarray}
T_1 &= & t_1+t_1G_0T_2,  \\
T_2 &= & t_2+t_2G_0T_1,   \\
T   &= & T_1+T_2, \label{fcaamp}
\end{eqnarray}
where the amplitudes $t_1$ and $t_2$ represent the unitary
scattering amplitudes with coupled channels for the interactions of
particle 3 with particle 1 and 2, respectively. The function $G_0$
in the above equations is the propagator for the particle 3 between
the particle 1 and 2 components of the cluster, which we will
discuss below.

We will calculate the total scattering amplitude $T$ in the low
energy regime, close to the threshold of the $\eta \bar{K} K^*$
systme or below, where the FCA is a good approximation, then the
on-shell approximation for the three particles are also used.

Following the field normalization of
Refs.~\cite{Roca:2010tf,YamagataSekihara:2010qk}, we can write down
the $S$-matrix for the single scattering term
[Fig.~\ref{fig:fca-diagram}(a) and 1(e)] as~\footnote{In principle,
there are two parts of the $S$-matrix: non-interaction part and
interaction part. Here, we consider only the interesting part of the
$S$-matrix that is the part due to interactions.}
\begin{eqnarray}
S^{(1)}=&S_1^{(1)}+S_2^{(1)} =
\frac{(2\pi)^4}{\mathcal{V}^2}\delta^4(k_3+k_{\rm{cls}}-k_3'-k_{\rm{cls}}')
\times
\nonumber \\
&\frac{1}{\sqrt{2w_3}}\frac{1}{\sqrt{2w_3'}}
(\frac{-it_1}{\sqrt{2w_1}} \frac{1}{\sqrt{2w_1'}} +
\frac{-it_2}{\sqrt{2w_2}} \frac{1}{\sqrt{2w_2'}}), \label{single}
\end{eqnarray}
where $\mathcal{V}$ stands for the volume of a box in which the
states are normalized to unity, while the momentum $k(k')$ and the
on-shell energy $w(w')$ refer to the initial (final) particles,
respectively.

The double scattering contributions are from
Fig.~\ref{fig:fca-diagram}(b) and 1(f). The expression for the
$S$-matrix for the double scattering [$S_2^{(2)}=S_1^{(2)}$] is
given by
\begin{align}
S^{(2)}=&-it_1t_2\frac{(2\pi)^4}{\mathcal{V}^2} \delta^4(k_3+k_{\rm{cls}}-k_3'-k_{\rm{cls}}') \nonumber \\
 &\times\frac{1}{\sqrt{2w_3}} \frac{1}{\sqrt{2w_3'}} \frac{1}{\sqrt{2w_1}} \frac{1}{\sqrt{2w_1'}} \frac{1}{\sqrt{2w_2}} \frac{1}{\sqrt{2w_2'}} \nonumber \\
& \times \int \frac{d^3q}{(2\pi)^3} F_{\rm{cls}}(q)
\frac{1}{{q^0}^2-|{\vec{q}}|^2-m_3^2+i \epsilon},\label{doble}
\end{align}
where the $F_{\rm{cls}}(q)$ is the form factor of the cluster which
is a bound state of particles 1 and 2. The information on the bound
state is encoded in the form factor $F_{\rm{cls}}(q)$ appearing in
Eq.~(\ref{doble}), which represents essentially the Fourier
transform of the cluster wave function. The variable $q^0$ is the
energy carried by the particle 3 in the center of mass frame of the
particle 3 and the cluster, which is given by
\begin{eqnarray}\label{16}
q^0(s)=\frac{s+m_3^2-m_{\rm{cls}}^2}{2\sqrt{s}},
\end{eqnarray}
with $s$ the invariant mass squared of the $\eta \bar{K} K^*$
system.

For the form factor $F_{\rm{cls}}(q)$, We take the following form
only for $s$-wave bound states, as it was discussed in
Refs.~\cite{Roca:2010tf,YamagataSekihara:2010qk,YamagataSekihara:2010pj}:
\begin{align}
F_{\rm{cls}}(q)=&\frac{1}{N}\int_{\vert \vec{p} \vert < \Lambda,\vert \vec{p}-\vec{q} \vert < \Lambda} d^3 \vec{p} ~ \frac{1}{2w_1(\vec{p})}\frac{1}{2w_2(\vec{p})} \nonumber \\
&\times\frac{1}{m_{\rm{cls}}-w_1(\vec{p})-w_2(\vec{p})} \frac{1}{2w_1(\vec{p}-\vec{q})}\frac{1}{2w_2(\vec{p}-\vec{q})} \nonumber \\
&\times\frac{1}{m_{\rm{cls}}-w_1(\vec{p}-\vec{q})-w_2(\vec{p}-\vec{q})},
\label{form}
\end{align}
where the normalization factor $N$ is
\begin{eqnarray}
N=\int_{\vert \vec{p} \vert < \Lambda} d^3 \vec{p} ~ \Big(
\frac{1}{2w_1(\vec{p})}\frac{1}{2w_2(\vec{p})}
\frac{1}{m_{\rm{cls}}-w_1(\vec{p})-w_2(\vec{p})}\Big)^2, \nonumber
\end{eqnarray}
with $m_{\rm{cls}}$ the mass of the cluster. Note that the width of
$K^*$ should be also included in the
$F_{\rm{cls}}(q)$~\cite{Xie:2011uw}. However, as will see later, the
masses of $f_1(1285)$ and $K_1(1270)$ are below the threshold of
$\bar{K}K^*$ and $\eta K^*$, the effect of the width of $K^*$ is
small and can be neglected.

Similarly, the full $S$ matrix for the scattering of particle 3 with
the cluster will be given by
\begin{align}
S=& -iT\frac{(2\pi)^4}{\mathcal{V}^2}\delta^4(k_3+k_{\rm{cls}}-k_3'-k_{\rm{cls}}')\nonumber \\
& \frac{1}{\sqrt{2w_3}} \frac{1}{\sqrt{2w_3'}}
\frac{1}{\sqrt{2w_{\rm{cls}}}}
\frac{1}{\sqrt{2w_{\rm{cls}}'}}.\label{total}
\end{align}

By comparing Eqs.~(\ref{single}), (\ref{doble}), and (\ref{total}), we see that we have to give a wight to $t_1$ and $t_2$ such that Eqs.~(\ref{single}) and (\ref{doble}) the weight factors that appear in Eq.~(\ref{total}). This is achieved by,

\begin{align}
\tilde{t}_1=t_1\sqrt{\frac{2w_{\rm{cls}}}{2w_1}}\sqrt{\frac{2w_{\rm{cls}}'}{2w_1'}},
~~~~~~
\tilde{t}_2=t_2\sqrt{\frac{2w_{\rm{cls}}}{2w_2}}\sqrt{\frac{2w_{\rm{cls}}'}{2w_2'}}.
\nonumber
\end{align}

Then, one can quickly solve Eqs.~(\ref{fcaamp}) and obtain
\begin{align}
T=\frac{\tilde{t}_1+\tilde{t}_2+2\tilde{t}_1\tilde{t}_2G_0}{1-\tilde{t}_1\tilde{t}_2G_0^2},
\end{align}
where $G_0$ depends on the invariant mass of the $\eta \bar{K}K^*$
system and it is given by
\begin{eqnarray}
G_0(s)=\frac{1}{2m_{\rm{cls}}}\int \frac{d^3q}{(2\pi)^3}
\frac{F_{\rm{cls}}(q)}{{q^0}^2 - |\vec{q}~|^2-m_3^2+i \epsilon} .
\end{eqnarray}

\subsection{Single scattering contribution}

On the other hand, it is worth noting that the argument of the total
scattering amplitude $T$ is regarded as a function of the total
invariant mass $\sqrt{s}$ of the three body system, while the
arguments of two body scattering amplitudes $t_1$ and $t_2$ depend
on the two body invariant mass $\sqrt{s_1}$ and $\sqrt{s_2}$, respectively. The
$s_1$ and $s_2$ are the invariant masses squared of the external
particle $3$ with momentum $k_3$ and particle 1 (2) inside the
cluster with momentum $k_1$ ($k_2$), which are given by

\begin{align}
s_1=m_3^2+m_1^2+\frac{(s-m_3^2-m_{ \rm{\rm{cls}}}^2)(m_{\rm{cls}}^2+m_1^2-m_2^2)}{2m_{\rm{cls}}^2}, \nonumber\\
s_2=m_3^2+m_2^2+\frac{(s-m_3^2-m_{\rm{cls}}^2)(m_{\rm{cls}}^2+m_2^2-m_1^2)}{2m_{\rm{cls}}^2},
\nonumber
\end{align}
where $m_l$ $(l=1,2,3)$ are the masses of the corresponding
particles in the three body system.

It is worth mentioning that in order to evaluate the two body
scattering amplitudes $t_1$ and $t_2$, the isospin of the cluster
should be considered. For the case of
$\eta$-$(\bar{K}K^*)_{f_1(1285)}$ system, the cluster of
$\bar{K}K^*$ has isospin $I_{\bar{K}K^*}=0$. Therefore, we have

\begin{align}
\vert \bar{K} K^* \rangle_{I=0}=\frac{1}{\sqrt{2}}\vert (\frac{1}{2}
,-\frac{1}{2}) \rangle-\frac{1}{\sqrt{2}}\vert (-\frac{1}{2}
,\frac{1}{2} )\rangle,
\end{align}
where the kets in the right-hand side indicate the $I_z$ components
of the particles $\bar{K}$ and $K^*$, $\vert (I_z^{\bar{K}}
,I_z^{K^*} )\rangle$. For the case of the total isospin
$I_{\eta(\bar{K}K^*)}=0$, the single scattering amplitude is written
as~\cite{Zhang:2016bmy}
\begin{widetext}
\begin{align}
\langle \eta(\bar{K}K^*)\vert t \vert \eta(\bar{K}K^*)
\rangle=&\Big(   \langle0 0 \vert \otimes    \frac{1}{\sqrt{2}}\Big(
\Big\langle (\frac{1}{2} ,-\frac{1}{2}) \Big\vert - \Big\langle
(-\frac{1}{2} ,\frac{1}{2})\Big \vert \Big) \Big) (t_{31}+t_{32})
\Big( \vert 0 0 \rangle \otimes
\frac{1}{\sqrt{2}}\Big( \Big\vert (\frac{1}{2} ,-\frac{1}{2}) \Big\rangle - \Big\vert (-\frac{1}{2} ,\frac{1}{2} )\Big\rangle \Big) \Big) \nonumber \\
=&\Big( \frac{1}{\sqrt{2}} \Big\langle
(\frac{1}{2}\frac{1}{2},-\frac{1}{2}) \Big\vert -
\frac{1}{\sqrt{2}}\Big\langle
(\frac{1}{2}\!\!-\!\!\frac{1}{2},\frac{1}{2})\Big \vert \Big) t_{31}
\Big( \frac{1}{\sqrt{2}}\Big \vert
(\frac{1}{2}\frac{1}{2},-\frac{1}{2})\Big \rangle -
\frac{1}{\sqrt{2}} \Big\vert
(\frac{1}{2}\!\!-\!\!\frac{1}{2},\frac{1}{2})\Big \rangle
 \Big) \nonumber \\
&+\Big( \frac{1}{\sqrt{2}}\Big\langle
(\frac{1}{2}\frac{1}{2},-\frac{1}{2}) \Big\vert -
\frac{1}{\sqrt{2}}\Big\langle
(\frac{1}{2}\!\!-\!\!\frac{1}{2},\frac{1}{2}) \Big\vert \Big) t_{32}
\Big( \frac{1}{\sqrt{2}}\Big \vert (\frac{1}{2}
\!\!-\!\!\frac{1}{2},\frac{1}{2}) \Big\rangle - \frac{1}{\sqrt{2}}
\Big\vert (\frac{1}{2}\frac{1}{2},-\frac{1}{2}) \Big\rangle \Big) ,
\end{align}
\end{widetext}
where the notation followed in the last term for the states is
$\vert(I_{\eta \bar{K}}I_{\eta \bar{K}}^z,I_{K^*}^z)\rangle$ for
$t_{31}$, while $\vert(I_{\eta K^*}I_{\eta
K^*}^z,I_{\bar{K}}^z)\rangle$ for $t_{32}$. This leads to the
following amplitudes for the single scattering contribution
[Fig.~\ref{fig:fca-diagram}(a) and 1(e)] in the
$\eta$-$(\bar{K}K^*)_{f_1(1285)}$ system,

\begin{align}
t_1=t_{\eta \bar{K} \to \eta \bar{K}}^{I=1/2},   \quad t_2=t_{\eta
K^* \to \eta K^*}^{I=1/2}.
\end{align}

In the $\bar{K}$-$(\eta K^*)_{K_1(1270)}$ system, the cluster $\eta
K^*$ can only have isospin $I_{\eta K^*}=1/2$. Therefore, for the
total isospin $I_{\bar{K}(\eta K^*)}=0$, the scattering amplitude is
written as~\cite{Zhang:2016bmy}
\begin{widetext}
\begin{align}
& \langle \bar{K}(\eta K^*)\vert t \vert \bar{K}(\eta K^*) \rangle =
\nonumber \\
& \frac{1}{\sqrt{2}} \Big(   \langle\frac{1}{2} \frac{1}{2} \vert
\otimes  \Big\langle (\frac{1}{2} ,-\frac{1}{2}) \Big\vert -
\langle\frac{1}{2} \!\!-\!\!\frac{1}{2} \vert \otimes \Big\langle
(\frac{1}{2} ,\frac{1}{2})\Big \vert  \Big) (t_{31}+t_{32})
\frac{1}{\sqrt{2}} \Big( \vert \frac{1}{2} \frac{1}{2} \rangle
\otimes
 \Big\vert (\frac{1}{2} ,-\frac{1}{2}) \Big\rangle -  \vert \frac{1}{2} \!\!-\!\!\frac{1}{2} \rangle \otimes \Big\vert (\frac{1}{2} ,\frac{1}{2} )\Big\rangle \Big) \nonumber \\
=&\Big( \frac{1}{\sqrt{2}} \Big\langle
(\frac{1}{2}\frac{1}{2},-\frac{1}{2}) \Big\vert -
\frac{1}{\sqrt{2}}\Big\langle
(\frac{1}{2}\!\!-\!\!\frac{1}{2},\frac{1}{2})\Big \vert \Big) t_{31}
\Big( \frac{1}{\sqrt{2}}\Big \vert
(\frac{1}{2}\frac{1}{2},-\frac{1}{2})\Big \rangle -
\frac{1}{\sqrt{2}} \Big\vert
(\frac{1}{2}\!\!-\!\!\frac{1}{2},\frac{1}{2})\Big \rangle
 \Big) \nonumber \\
&+\frac{1}{\sqrt{2}}\Big( \frac{1}{\sqrt{2}}\Big\langle (00,0) +
\frac{1}{\sqrt{2}}\Big\langle (00,0) \Big\vert \Big)  t_{32}
\frac{1}{\sqrt{2}}\Big( \frac{1}{\sqrt{2}} \Big\vert (00,0)\Big
\rangle + \frac{1}{\sqrt{2}} \Big\vert (00,0) \Big\rangle \Big).
\end{align}
\end{widetext}
This leads to the following amplitudes for the single scattering
contribution in the $\bar{K}$-$(\eta K^*)_{K_1(1270)}$ system,

\begin{align}
t_1=t_{\bar{K} \eta \to \bar{K} \eta}^{I=1/2}, \quad t_2=t_{\bar{K}
K^* \to \bar{K} K^*}^{I=0}.
\end{align}
We see that only the $\bar{K}K^* \to \bar{K}K^*$ in $I=0$ transition
has contribution, since we want to make the total isospin
$I_{\bar{K}(\eta K^*)}=0$, and the $\eta$ meson has isospin zero.

\subsection{Unitarized $\eta K^*$ and $\bar{K} K^*$ interactions}

The important ingredients in the calculation of the total scattering
amplitude for the $\eta \bar{K} K^*$ system using the FCA are the
two body $\eta K$, $\eta K^*$, and $\bar{K} K^*$ unitarized $s$ wave
interactions from the chiral unitary approach. These two body
scattering amplitudes are studied with the dimensional
regularization procedure, and they depend on the subtraction
constants $a_{\eta K}$, $a_{\eta K^*}$ and $a_{\bar{K}K^*}$, and
also the regularization scale $\mu$. Note that there is only one
parameter for the dimensional regularization procedure, since any
change in $\mu$ is reabsorbed by a change in $a(\mu)$ through
$a(\mu') - a(\mu) = {\rm ln}(\frac{\mu'^2}{\mu^2})$ so that the
scattering amplitude is scale independent. In this work, we take
these parameters as used in
Refs.~\cite{Roca:2005nm,Geng:2006yb,Guo:2005wp}: $a_{\eta K}=-1.38$
and $\mu=m_K$ for $I_{\eta K}=1/2$; $a_{\eta K^*}=-1.85$ and
$\mu=1000$ MeV for $I_{\eta K^*}=1/2$; $a_{\bar{K}K^*}=-1.85$ and
$\mu=1000$ MeV for $I_{\bar{K}K^*}=0$. With those parameters, one
can get the mass of $f_1(1285)$ and $K_1(1270)$ at their estimated
values.~\footnote{More details for the two body scattering can be
found in Refs.~\cite{Roca:2005nm,Geng:2006yb,Guo:2005wp}.}

\begin{figure}[htbp]
\begin{center}
\includegraphics [scale=0.6] {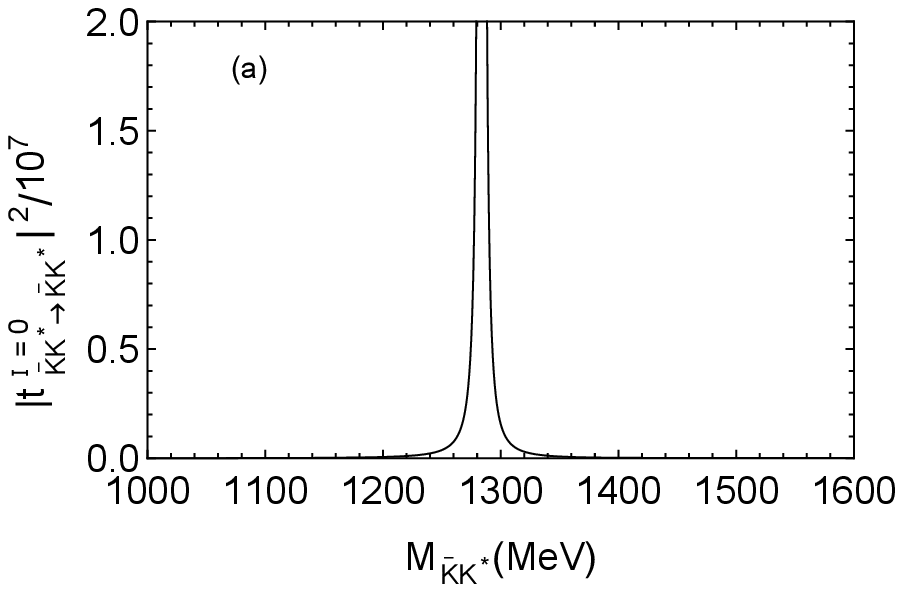}\vspace{0.5cm}
\includegraphics [scale=0.6] {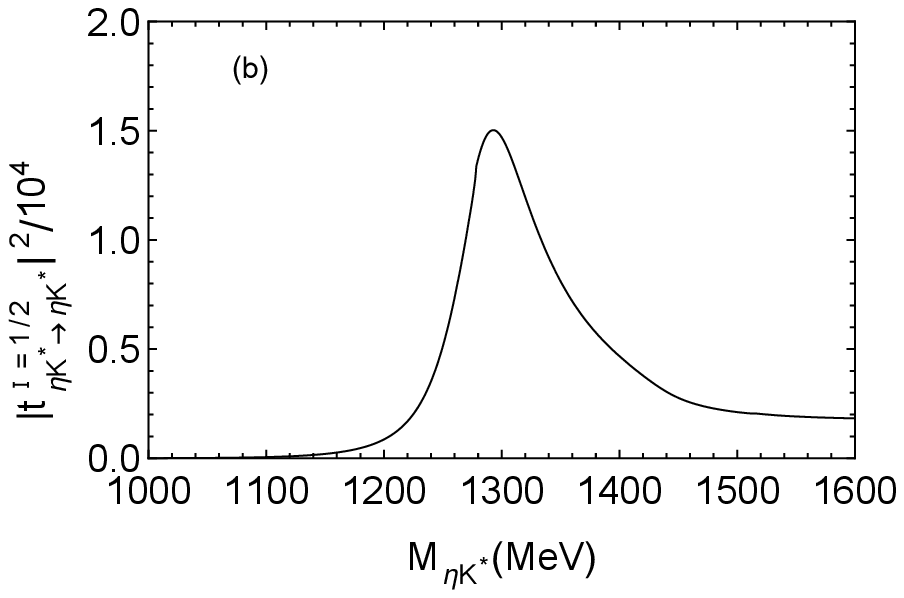}
\caption{(a) Modulus squared of $t_{\bar{K} K^* \to \bar{K}
K^*}^{I=0}$ as a function of the invariant mass $M_{\bar{K}K^*}$ of
the $\bar{K}K^*$ subsystem. (b) Modulus squared of $t_{\eta K^* \to
\eta K^*}^{I=1/2}$ as a function of the invariant mass $M_{\eta
K^*}$ of the $\eta K^*$ subsystem.} \label{fig:tsquare}
\end{center}
\end{figure}

In Figs.~\ref{fig:tsquare} (a) and (b) we show the numerical results
for $|t_{\bar{K} K^* \to \bar{K} K^*}^{I=0}|^2$ and $|t_{\eta K^*
\to \eta K^*}^{I=1/2}|^2$, respectively, from where we see clear
peaks for $f_1(1285)$ and $K_1(1270)$ states.

\subsection{Form factors $F_{\rm{cls}}(q)$ and propagator $G_0(s)$}

To connect with the dimensional regularization procedure we choose
the cutoff $\Lambda$ such that the value of the two body loop
function at threshold coincides in both methods. Thus we take
$\Lambda=990$ MeV such that the $f_1(1285)$ is obtained in
Refs.~\cite{Xie:2015lta,Aceti:2015pma}, while for $K_1(1270)$ we
take $\Lambda= 1000$ MeV. The cut-off is tuned to get a pole at
$1288- i74$ for the $K_1(1270)$ state.

\begin{figure}[htbp]
\begin{center}
\includegraphics [scale=0.45] {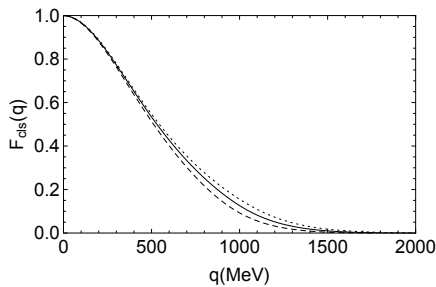}
\caption{Forms factor of Eq.~(\ref{form}) in terms of $q =
|\vec{q}~|$ with cut-off $\Lambda=890$ (dashed), $990$ (solid), and
$1090$ MeV (dotted) for $f_1(1285)$ as a $\bar{K}K^*$ bound state.}
\label{fig:factor-1285}
\end{center}
\end{figure}

\begin{figure}[htbp]
\begin{center}
\includegraphics [scale=0.45] {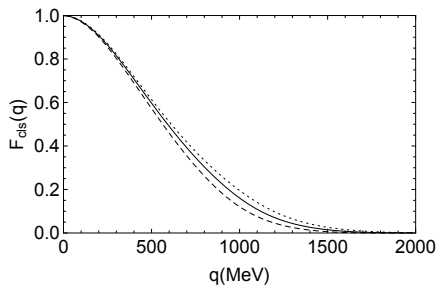}
\caption{As in Fig.~\ref{fig:factor-1285} but for the case of
$K_1(1270)$ as a $\eta K^*$ bound state. The dashed, solid and
dotted curves are obtained with $\Lambda=900$, 1000, and $1100$ MeV,
respectively.} \label{fig:factor-1270}
\end{center}
\end{figure}

In Figs.~\ref{fig:factor-1285} and \ref{fig:factor-1270} we show the
respective form factors for the $f_1(1285)$ and $K_1(1270)$,
respectively, where we take $m_{\rm{cls}}=1281.3$ MeV for
$f_1(1285)$ and 1284 MeV for $K_1(1270)$ as obtained in
Ref.~\cite{Geng:2006yb}. In the FCA, we keep the wave function of
the cluster unchanged by the presence of the third particle. In
order to estimate uncertainties of the FCA due to this ``frozen"
condition we admit that the wave function of the cluster could be
modified by the presence of the third particle. For doing this, we
perform calculations with different cut offs. These results shown in
Figs.~\ref{fig:factor-1285} and \ref{fig:factor-1270} are obtained
with cutoff $\Lambda = 890$, $990$ and $1090$ MeV for the case of
$f_1(1285)$, while for the case of $K_1(1270)$, we take $\Lambda =
900$, $1000$ and $1100$ MeV.

\begin{figure}[htbp]
\begin{center}
\includegraphics [scale=0.5] {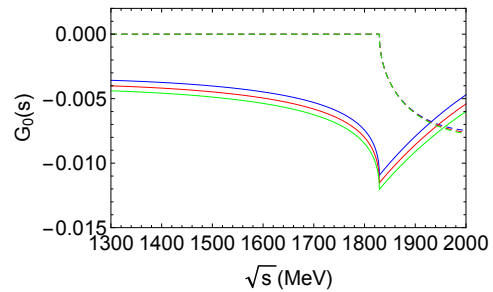}
\caption{Real (solid curve) and imaginary (dashed curve) parts of
the $G_0$ function for the $\eta$-$(\bar{K}K^*)_{f_1(1285)}$ system.
These results are obtained with cutoff $\Lambda = 890$ (blue), $990$
(red) and $1090$ MeV (green).} \label{fig:propa1}
\end{center}
\end{figure}

Next, in Fig.~\ref{fig:propa1}, we show the real (solid curves) and
imaginary (dashed curves) parts of the $G_0$ as a function of the
invariant mass for the $\eta$-$(\bar{K}K^*)_{f_1(1285)}$ system with
cutoff $\Lambda = 890$, $990$ and $1090$ MeV.

\begin{figure}[htbp]
\begin{center}
\includegraphics [scale=0.5] {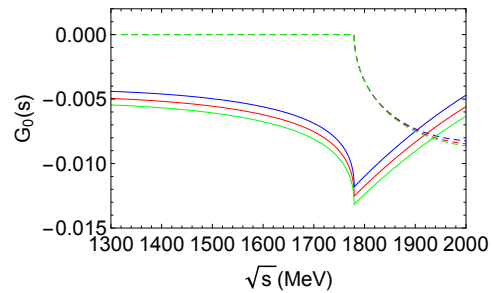}
\caption{Real (solid curve) and imaginary (dashed curve) parts of
the $G_0$ function for the $\bar{K}$-$(\eta K^*)_{K_1(1270)}$
system. These results are obtained with cutoff $\Lambda = 900$
(blue), $1000$ (red) and $1100$ MeV (green).} \label{fig:propa2}
\end{center}
\end{figure}

The results of the $G_0$ function for the $\bar{K}$-$(\eta
K^*)_{K_1(1270)}$ system are shown in Fig.~\ref{fig:propa2}, we show
the real (solid curves) and imaginary (dashed curves) parts ,
$\Lambda = 900$, $1000$ and $1100$ MeV.

From Figs.~\ref{fig:propa1} and \ref{fig:propa2} one can see that
the imaginary part of $G_0(s)$ is not sensitive the values of
cutoff, while the real part changes a bit with the changing of
cutoff.

\section{Numerical results and discussion}

For the numerical evaluation of the three body amplitude, we shall
need the calculation of two body interaction amplitudes of $\eta
\bar{K}$, $\eta K^*$, and $\bar{K} K^*$, which were investigated by
the chiral dynamics and unitary coupled channels approach in
Refs.~\cite{Roca:2005nm,Geng:2006yb,Guo:2005wp}. Then we calculate
the total scattering amplitude $T$ and associate the peaks or bumps
in the modulus squared $\vert T \vert ^2$ to resonances.

\begin{figure}[htbp]
\begin{center}
\includegraphics [scale=0.45] {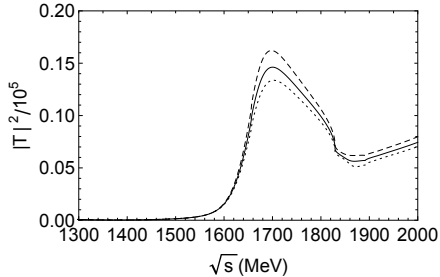}
\caption{Modulus squared of the total amplitudes $T$ for the
$\eta$-$(\bar{K}K^*)_{f_1(1285)}$ system. The dashed, solid and
dotted curves are obtained with $\Lambda=890$, 990, and $1090$ MeV,
respectively.} \label{fig:numto1}
\end{center}
\end{figure}

In Fig.~\ref{fig:numto1} we show the modulus squared $|T|^2$ for the
$\eta$-$(\bar{K}K^*)_{f_1(1285)}$ scattering with total isospin
$I=0$. One can see that there is a clear bump structure which is
below the $\eta{f_1(1285)}$ threshold with mass around 1700 MeV and
width about 180 MeV. Furthermore, taking $\sqrt{s}=1700$ MeV, we get
$\sqrt{s_1}=927$ MeV and $\sqrt{s_2}=1315$ MeV. At this energy
point, the interactions of $\eta \bar{K}$ and $\eta K^*$ are strong.

\begin{figure}[htbp]
\begin{center}
\includegraphics [scale=0.45] {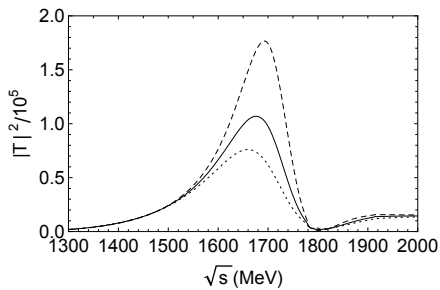}
\caption{Modulus squared of the total amplitudes $T$ for the
$\bar{K}$-$(\eta K^*)_{K_1(1270)}$ system. The dashed, solid and
dotted curves are obtained with $\Lambda=900$, 1000, and $1100$ MeV,
respectively.} \label{fig:numto2}
\end{center}
\end{figure}

In Fig.~\ref{fig:numto2}, we show the results of the $\vert T \vert
^2$ for the $\bar{K}$-$(\eta K^*)_{K_1(1270)}$ system. A strong
resonant structure around 1680 MeV with a width about 160 MeV shows
up, which indicates that a
$\bar{K}$-$(\eta K^*)_{K_1(1270)}$ state can be formed. The mass of
the state is below the $\bar{K}$ and ${K_1(1270)}$ mass threshold.
The strength of $|T|^2$ at the peak is much larger than that of
Fig.~\ref{fig:numto1} for the case of $\eta f_1(1285) \to \eta
f_1(1285)$ scattering. Thus, it is clear that the preferred
configuration is $\bar{K} K_1(1270)$. However, the $\bar{K}$ will
keep interacting with the $K^*$ and sometimes can also make
$f_1(1285)$.

From Figs.~\ref{fig:numto1} and \ref{fig:numto2}, it can be seen
that the peak positions and widths of
$\eta$-$(\bar{K}K^*)_{f_1(1285)}$ and $\bar{K}$-$(\eta
K^*)_{K_1(1270)}$ systems are quite stable with small variations of
the cutoff $\Lambda$ parameter.~\footnote{The strength of $|T|^2$
changes a bit with the changing of the cutoff parameter.} This gives
confidence that the $\eta$-$(\bar{K}K^*)_{f_1(1285)}$ and
$\bar{K}$-$(\eta K^*)_{K_1(1270)}$ bound states can be formed. In
fact, the $\eta f_1(1285)$ configuration would mix with the
$\bar{K}K_1(1270)$ configuration. However, since the strength of the
$\bar{K}$-$(\eta K^*)_{K_1(1270)}$ scattering is much larger than
the one of $\eta$-$(\bar{K}K^*)_{f_1(1285)}$ scattering, the
interference between the two configurations should be small, and
both configurations peak around the similar energy, thus, it is
expected that the peak around this energy of any mixture of the
state is guaranteed.

The $\eta \bar{K} K^*$ bound state with quantum numbers $ I(J^P) = 0
(1^-)$ has a dominant $\bar{K}K_1(1270)$ component. Since
$K_1(1270)$ decays into $K \pi \pi $ mostly~[49], the dominant decay
mode of the proposed state should be $\bar{K}K\pi \pi$, and one hope
that the future experimental measurements can test our model
calculations.

We should mention that there are two $K_1(1270)$ states are obtained
in Ref.~\cite{Geng:2006yb}. The higher one with mass 1284 MeV
couples more strongly to the $\eta K^*$ and $K \rho$ channels, while
the lower one with mass 1195 MeV mainly couples to the $\pi K^*$
channel, and it couples to the $\eta K^*$ very weak. Thus, one can
expect that the lower $K_1(1270)$ of Ref.~\cite{Geng:2006yb} will
not affect the calculations here.

\section{Summary}

In this work, we have used the FCA to the Faddeev equations in order
to look for possible $I^G(J^{PC})=0^+(1^{-+})$ exotic states
generated from $\eta\bar{K}K^*$ three body interactions. We first
select a cluster of $\bar{K}K^*$, which is known to generate the
$f_1(1285)$ in $I=0$, and then let the $\eta$ meson interact with
$\bar{K}$ and $K^*$. In the modulus squared of
$\eta$-$(\bar{K}K^*)_{f_1(1285)}$ scattering amplitude, we find
evidence of a bound state below the $\eta{f_1(1285)}$ threshold with
mass around 1700 MeV and width about 100 MeV. In the case of
$\bar{K}$ scattering with the cluster $\eta K^*$, which is shown to
generate the $K_1(1270)$ in $I=1/2$, we obtain a bound state
$I(J^{P})=0(1^{-})$ just below the $\bar{K}{K_1(1270)}$ threshold
with mass around 1680 MeV and width about 160 MeV. In addition, the
simplicity of the present approach also allows for a transparent
interpretation of the results, which is not easy to see when one
uses the full Faddeev equations. In the present study, it is easy to
know that the $\bar{K}K_1(1270)$ is the dominant one, where the
$\bar{K}K^*$ subsystem can still couple to the $f_1(1285)$
resonance. Yet, one may think that we should rely on the full
Faddeev calculations where all the scattering processes can be
summed up to infinite order, for example, as pointed in the
Refs.~\cite{Miyagawa:2012xz,Miyagawa:2018xge} for the study of $K^-
d \to \pi \Sigma n$ reaction. Such calculations are welcome and we
intend to address this issue in a future study.

The predictions of existence of possible exotic states have been
made within the framework of flux tube model~\cite{Page:1998gz},
Lattice QCD \cite{Bernard:2003jd} and QCD sum
rule~\cite{Chen:2008ne}. The results obtained here provide a
different theoretical approach for a devoted investigation of these
exotic states.

\section*{Acknowledgments}

We would like to thank Prof. Li-Sheng Geng for useful discussions.
This work is partly supported by the National Natural Science
Foundation of China under Grant Nos.~11735003, 1191101015, 11475227,
and by the Youth Innovation Promotion Association CAS (No.~2016367).


\begin{thebibliography}{99}

\bibitem{Liu:2019zoy}
  Y.~R.~Liu, H.~X.~Chen, W.~Chen, X.~Liu and S.~L.~Zhu,
  Prog.\ Part.\ Nucl.\ Phys.\  {\bf 107}, 237 (2019).

\bibitem{Guo:2017jvc}
  F.~K.~Guo, C.~Hanhart, U.~G.~Mei\ss ner, Q.~Wang, Q.~Zhao and B.~S.~Zou,
  Rev.\ Mod.\ Phys.\  {\bf 90}, 015004 (2018).


\bibitem{Chen:2016spr}
  H.~X.~Chen, W.~Chen, X.~Liu, Y.~R.~Liu and S.~L.~Zhu,
  Rept.\ Prog.\ Phys.\  {\bf 80}, 076201 (2017).


\bibitem{Amsler:2004ps}
  C.~Amsler and N.~A.~Tornqvist,
  Phys.\ Rept.\  {\bf 389}, 61 (2004).


\bibitem{Tanabashi:2018oca}
  M.~Tanabashi {\it et al.} [Particle Data Group],
  Phys.\ Rev.\ D {\bf 98}, 030001 (2018).


\bibitem{Isgur:1984bm}
  N.~Isgur and J.~E.~Paton,
  Phys.\ Rev.\ D {\bf 31}, 2910 (1985).


\bibitem{Close:1994hc}
  F.~E.~Close and P.~R.~Page,
  Nucl.\ Phys.\ B {\bf 443}, 233 (1995).


\bibitem{Page:1998gz}
  P.~R.~Page, E.~S.~Swanson and A.~P.~Szczepaniak,
  Phys.\ Rev.\ D {\bf 59}, 034016 (1999).


\bibitem{Kim:2008qh}
  H.~C.~Kim and Y.~Kim,
  JHEP {\bf 0901}, 034 (2009).


\bibitem{Bellantuono:2014lra}
  L.~Bellantuono, P.~Colangelo and F.~Giannuzzi,
  Eur.\ Phys.\ J.\ C {\bf 74}, 2830 (2014).


\bibitem{Dudek:2010wm}
  J.~J.~Dudek, R.~G.~Edwards, M.~J.~Peardon, D.~G.~Richards and C.~E.~Thomas,
  Phys.\ Rev.\ D {\bf 82}, 034508 (2010).


\bibitem{Dudek:2009qf}
  J.~J.~Dudek, R.~G.~Edwards, M.~J.~Peardon, D.~G.~Richards and C.~E.~Thomas,
  Phys.\ Rev.\ Lett.\  {\bf 103}, 262001 (2009).


\bibitem{Bernard:2003jd}
  C.~Bernard {\it et al.},
  Phys.\ Rev.\ D {\bf 68}, 074505 (2003).

\bibitem{Huang:2017pzh}
  Z.~R.~Huang, H.~Y.~Jin, T.~G.~Steele and Z.~F.~Zhang,
  Nucl.\ Part.\ Phys.\ Proc.\  {\bf 294-296}, 113 (2018).

\bibitem{Huang:2016upt}
  Z.~R.~Huang, H.~Y.~Jin, T.~G.~Steele and Z.~F.~Zhang,
  Phys.\ Rev.\ D {\bf 94}, 054037 (2016).

\bibitem{Huang:2014hya}
  Z.~R.~Huang, H.~Y.~Jin and Z.~F.~Zhang,
  JHEP {\bf 1504}, 004 (2015).

\bibitem{Chen:2010ic}
  H.~X.~Chen, Z.~X.~Cai, P.~Z.~Huang and S.~L.~Zhu,
  Phys.\ Rev.\ D {\bf 83}, 014006 (2011).

\bibitem{Chen:2008qw}
  H.~X.~Chen, A.~Hosaka and S.~L.~Zhu,
  Phys.\ Rev.\ D {\bf 78}, 054017 (2008).

\bibitem{Narison:2009vj}
  S.~Narison,
  Phys.\ Lett.\ B {\bf 675}, 319 (2009).

\bibitem{Zhang:2016bmy}
  X.~Zhang, J.~J.~Xie and X.~Chen,
  Phys.\ Rev.\ D {\bf 95}, 056014 (2017).


\bibitem{Roca:2005nm}
  L.~Roca, E.~Oset and J.~Singh,
  Phys.\ Rev.\ D {\bf 72}, 014002 (2005).


\bibitem{Lutz:2003fm}
  M.~F.~M.~Lutz and E.~E.~Kolomeitsev,
  Nucl.\ Phys.\ A {\bf 730}, 392 (2004).


\bibitem{Chen:2008ne}
  H.~X.~Chen, A.~Hosaka and S.~L.~Zhu,
  Phys.\ Rev.\ D {\bf 78}, 117502 (2008).


\bibitem{Gal:2006cw}
  A.~Gal,
  Int.\ J.\ Mod.\ Phys.\ A {\bf 22}, 226 (2007).

\bibitem{Barrett:1999cw}
  R.~C.~Barrett and A.~Deloff,
  Phys.\ Rev.\ C {\bf 60}, 025201 (1999).

\bibitem{Kamalov:2000iy}
  S.~S.~Kamalov, E.~Oset and A.~Ramos,
  Nucl.\ Phys.\ A {\bf 690}, 494 (2001).


\bibitem{Bayar:2011qj}
  M.~Bayar, J.~Yamagata-Sekihara and E.~Oset,
  Phys.\ Rev.\ C {\bf 84}, 015209 (2011).

\bibitem{Bayar:2012hn}
  M.~Bayar and E.~Oset,
  Phys.\ Rev.\ C {\bf 88}, 044003 (2013).

\bibitem{Ajimura:2018iyx}
  S.~Ajimura {\it et al.} [J-PAC E15 Collaboration],
  Phys.\ Lett.\ B {\bf 789}, 620 (2019).


\bibitem{Xie:2011uw}
  J.~J.~Xie, A.~Martinez Torres, E.~Oset and P.~Gonzalez,
  Phys.\ Rev.\ C {\bf 83}, 055204 (2011).


\bibitem{Xie:2010ig}
  J.~J.~Xie, A.~Martinez Torres and E.~Oset,
  Phys.\ Rev.\ C {\bf 83}, 065207 (2011).


\bibitem{Jido:2008kp}
  D.~Jido and Y.~Kanada-En'yo,
  Phys.\ Rev.\ C {\bf 78}, 035203 (2008).


\bibitem{MartinezTorres:2009cw}
  A.~Martinez Torres, K.~P.~Khemchandani, U.~G.~Meissner and E.~Oset,
  Eur.\ Phys.\ J.\ A {\bf 41}, 361 (2009).

\bibitem{MartinezTorres:2008kh}
  A.~Martinez Torres, K.~P.~Khemchandani and E.~Oset,
  Phys.\ Rev.\ C {\bf 79}, 065207 (2009).

\bibitem{MartinezTorres:2010zv}
  A.~Martinez Torres and D.~Jido,
  Phys.\ Rev.\ C {\bf 82}, 038202 (2010).

\bibitem{Ren:2018qhr}
  X.~L.~Ren and Z.~F.~Sun,
  Phys.\ Rev.\ D {\bf 99}, 094041 (2019).

\bibitem{Dias:2018iuy}
  J.~M.~Dias, L.~Roca and S.~Sakai,
  Phys.\ Rev.\ D {\bf 97}, 056019 (2018).

\bibitem{Bayar:2015zba}
  M.~Bayar, P.~Fernandez-Soler, Z.~F.~Sun and E.~Oset,
  Eur.\ Phys.\ J.\ A {\bf 52}, 106 (2016).

\bibitem{Bayar:2015oea}
  M.~Bayar, X.~L.~Ren and E.~Oset,
  Eur.\ Phys.\ J.\ A {\bf 51}, 61 (2015).

\bibitem{Xiao:2012dw}
  C.~W.~Xiao, M.~Bayar and E.~Oset,
  Phys.\ Rev.\ D {\bf 86}, 094019 (2012).

\bibitem{Debastiani:2017vhv}
  V.~R.~Debastiani, J.~M.~Dias and E.~Oset,
  Phys.\ Rev.\ D {\bf 96}, 016014 (2017).


\bibitem{Dias:2017miz}
  J.~M.~Dias, V.~R.~Debastiani, L.~Roca, S.~Sakai and E.~Oset,
  Phys.\ Rev.\ D {\bf 96}, 094007 (2017).

\bibitem{Ren:2018pcd}
  X.~L.~Ren, B.~B.~Malabarba, L.~S.~Geng, K.~P.~Khemchandani and A.~Mart¨ªnez Torres,
  Phys.\ Lett.\ B {\bf 785}, 112 (2018).

\bibitem{MartinezTorres:2018zbl}
  A.~Martinez Torres, K.~P.~Khemchandani and L.~S.~Geng,
  Phys.\ Rev.\ D {\bf 99}, 076017 (2019).

\bibitem{SanchezSanchez:2017xtl}
  M.~Sanchez Sanchez, L.~S.~Geng, J.~X.~Lu, T.~Hyodo and M.~P.~Valderrama,
  Phys.\ Rev.\ D {\bf 98}, 054001 (2018).

\bibitem{Wu:2019vsy}
  T.~W.~Wu, M.~Z.~Liu, L.~S.~Geng, E.~Hiyama and M.~P.~Valderrama,
  Phys.\ Rev.\ D {\bf 100}, 034029 (2019).

\bibitem{MartinezTorres:2010ax}
  A.~Martinez Torres, E.~J.~Garzon, E.~Oset and L.~R.~Dai,
  Phys.\ Rev.\ D {\bf 83}, 116002 (2011).

\bibitem{MartinezTorres:2008gy}
  A.~Martinez Torres, K.~P.~Khemchandani, L.~S.~Geng, M.~Napsuciale and E.~Oset,
  Phys.\ Rev.\ D {\bf 78}, 074031 (2008).

\bibitem{Geng:2006yb}
  L.~S.~Geng, E.~Oset, L.~Roca and J.~A.~Oller,
  Phys.\ Rev.\ D {\bf 75}, 014017 (2007).

\bibitem{Guo:2005wp}
  F.~K.~Guo, R.~G.~Ping, P.~N.~Shen, H.~C.~Chiang and B.~S.~Zou,
  Nucl.\ Phys.\ A {\bf 773}, 78 (2006).

\bibitem{Deloff:1999gc}
  A.~Deloff,
  Phys.\ Rev.\ C {\bf 61}, 024004 (2000).

\bibitem{Roca:2010tf}
  L.~Roca and E.~Oset,
  Phys.\ Rev.\ D {\bf 82}, 054013 (2010).

\bibitem{YamagataSekihara:2010qk}
  J.~Yamagata-Sekihara, L.~Roca and E.~Oset,
  Phys.\ Rev.\ D {\bf 82}, 094017 (2010)
  Erratum: [Phys.\ Rev.\ D {\bf 85}, 119905 (2012)].

\bibitem{YamagataSekihara:2010pj}
  J.~Yamagata-Sekihara, J.~Nieves and E.~Oset,
  Phys.\ Rev.\ D {\bf 83}, 014003 (2011).

\bibitem{Xie:2015lta}
  J.~J.~Xie and E.~Oset,
  Phys.\ Lett.\ B {\bf 753}, 591 (2016).

\bibitem{Aceti:2015pma}
  F.~Aceti, J.~J.~Xie and E.~Oset,
  Phys.\ Lett.\ B {\bf 750}, 609 (2015).

\bibitem{Miyagawa:2012xz}
  K.~Miyagawa and J.~Haidenbauer,
  Phys.\ Rev.\ C {\bf 85}, 065201 (2012).

\bibitem{Miyagawa:2018xge}
  K.~Miyagawa, J.~Haidenbauer and H.~Kamada,
  Phys.\ Rev.\ C {\bf 97}, 055209 (2018).

\end{thebibliography}
\end{document}